\documentclass{biophys-new}
\usepackage[utf8]{inputenc}
\usepackage{graphicx}
\usepackage[colorlinks,allcolors=cyan!70!black]{hyperref}

\usepackage{lipsum}

\newcommand{\Dcm}{D_{\rm CM}}
\newcommand{\DR}{D_{\rm g}}
\newcommand{\Ds}{D_{\rm self}}
\newcommand{\Dgrad}{\bar{D}}
\newcommand{\kT}{k_{\rm B}T}
\newcommand{\tenA}{\mathbf{A}}
\newcommand{\tenB}{\mathbf{B}}

\newcommand{\indA}{\mathrm{A}}
\newcommand{\indB}{\mathrm{B}}
\newcommand{\indD}{\mathrm{D}}

\newcommand{\vecf}{{\bf f}}
\newcommand{\vecF}{{\bf F}}
\newcommand{\vecr}{{\bf r}}

\newcommand{\vrho}{\boldsymbol{\rho}}
\newcommand{\vDelta}{\boldsymbol{\Delta}}
\newcommand{\const}{\frac{\kT}{4\pi\mu}}
\newcommand{\RG}{R_{\rm g}}
\newcommand{\Rcm}{\mathbf{R}_{\rm CM}}

\title{Persistent Collective Motion of a Dispersing Membrane Domain}
\runningtitle{Dispersion of Membrane Domain} 

\author[1]{Benjamin Sorkin}
\author[1,*]{Haim Diamant}
\runningauthor{Sorkin and Diamant} 

\affil[1]{Raymond and Beverly Sackler School of Chemistry, and Center for Physics and Chemistry of Living Systems, 
Tel Aviv University, Tel Aviv 6997801, Israel}

\corrauthor[*]{hdiamant@tau.ac.il}

\papertype{Article}

\begin{document}

\begin{frontmatter}

\begin{abstract} 
We study the Brownian motion of an assembly of mobile inclusions embedded in a fluid membrane. The motion includes the dispersal of the assembly, accompanied by the diffusion of its center of mass. Usually, the former process is much faster than the latter, since the diffusion coefficient of the center of mass is inversely proportional to the number of particles. However, in the case of  membrane inclusions, we find that the two processes occur on the same time scale, thus prolonging significantly the lifetime of the assembly as a collectively moving object. This effect is caused by the quasi-two-dimensional membrane flows, which couple the motions even of the most remote inclusions in the assembly. The same correlations also cause the diffusion coefficient of the center of mass to decay slowly with time, resulting in weak sub-diffusion. We confirm our analytical results by Brownian dynamics simulations with flow-mediated correlations. The effect reported here should have implications for the stability of nano-scale membrane heterogeneities.
\end{abstract}

\begin{sigstatement} 
Membrane heterogeneities on the scale of tens of nanometers  play a key role in many cellular functions. We show that once such a domain starts dispersing into the background lipid membrane, its collective random motion remains fast compared to the rate of dispersal, making the domain move as a compact object for times orders of magnitude longer than expected. The collective dynamics is caused by strong, long-ranged velocity correlations mediated by flows in the host fluid membrane. This fundamental flow-induced effect should be dominant at distances larger than the molecular scale, thus prolonging the stability and sustaining the collective motion of membrane nano-domains.  
\end{sigstatement}
\end{frontmatter}

\section*{Introduction}

Processes occurring on the cell membrane, such as signaling, adhesion, bridging, and membrane remodeling, usually involve the recruitment of specialized lipids and proteins into compact domains nanometers to tens of nanometers in size \cite{alberts2018molecular}. In addition, reports and arguments concerning the presence of so-called lipid rafts of different compositions have persisted for decades. The exact nature and stability of such 10-nanometers-scale dynamic domains have been the subject of intense research and debate \cite{goni2019rafts,raftreview2017}.

Due to the fluid nature of the lipid membrane, inclusions such as membrane proteins and small domains undergo random Brownian motion. The diffusion coefficient of a single inclusion has been studied extensively since the pioneering work of Saffman and Delbr\"uck (SD) \cite{SD75,saffman1976brownian}. The anomalous quasi-two-dimensional flows in a pristine membrane make the diffusion coefficient depend very weakly (logarithmically) on the particle's size. This holds so long as the particle is small compared to the so-called SD length, which is proportional to the ratio between membrane viscosity and the viscosity of the surrounding fluid. For a pristine membrane embedded in a fluid with the viscosity of water, the SD length is about half a micrometer, which is typically much larger than the radius of the inclusions mentioned above. These predictions have been confirmed in many experiments and simulations \cite{lyman2018dynamics,brown2011continuum,ramadurai2009lateral} but, at the same time, were found to be sensitive to details such as 
protein crowding~\cite{javanainen2017diffusion}, and the obstructed membrane structure in the vicinity of the inclusion \cite{naji2007corrections,ramadurai2009lateral}.

The SD model has been extended over the years to other scenarios, including objects whose size is comparable to and larger than the SD length \cite{hughes1981translational}, supported membranes \cite{stone1998hydrodynamics,seki2011diffusion}, and membranes containing immobile inclusions \cite{bussell1995effect,oppenheimer2011plane}. In the latter two cases the governing length scale is no longer the SD length but a smaller length dictated by the impurities (the substrate or immobile particles).

In recent years there has been a growing recognition that membrane inclusions do not move independently \cite{lyman2018dynamics,chein2019flow}. 
The same anomalous flows mentioned above cause the motions of lipids and proteins to be strongly correlated over large distances~\cite{falck2008lateral}. Up to a cutoff distance, determined by the SD length (or the smaller length in the non-pristine cases), the correlation decays only logarithmically with distance \cite{bussell1992resistivity,oppenheimer2009correlated,noruzifar2014calculating}. Even in the presence of impurities, the correlation beyond the cutoff length is not negligible but continues to decay as the inverse of the distance squared \cite{oppenheimer2010correlated,oppenheimer2011plane}. Such long-range correlations were measured in several synthetic membranes, and recently also in membranes of living cells \cite{chein2019flow}.

The dynamics of membranal assemblies is obviously complex and depends on various interactions and system parameters. In this work we isolate and consider a single effect, namely, the collective diffusion of the inclusions making the assembly under the flow-mediated correlations mentioned above. Unlike the direct interactions among inclusions (e.g., screened electrostatic interactions, van der Waals attraction, specific steric effects, etc.), it is a long-range effect which should dominate at separations larger than a few nanometers. This physical effect should be present in any particular scenario where an assembly of inclusions disperses over the fluid membrane. 
As a specific example, we mention the membrane fusion involved in cell infection by various membrane-enveloped viruses. In some viruses, this process was found to require a pH-induced, reversible, conformational changes of glycoproteins, which cause them to form a cluster of a few dozens of protein complexes \cite{kim2017mechanism}. The formation of this contact zone catalyzes the membrane fusion stage. Once the fusion process has completed, an opposite pH change can restore the original glycoproteins' conformations, thus enabling the dispersal of the cluster. 

When the diffusion of an assembly of mobile inclusions is concerned, there are several diffusion coefficients to consider: (a) the self-diffusion coefficient of a single isolated particle, $D_0$; (b) the self-diffusion coefficient of a single particle within the assembly, $\Ds$; (c) the diffusion coefficient of the assembly's center of mass (CM), $\Dcm$; and (d) the coefficient characterizing the dispersal of the assembly, i.e., the diffusive growth of its radius of gyration, $\DR$. The coefficients $D_0$ and $\Ds$ are measured, for example, using single-molecule tracking for an isolated membrane protein and a protein within an assembly, respectively \cite{lyman2018dynamics}. The coefficient $D_0$ can be measured also by fluorescence correlation spectroscopy (FCS) \cite{lippincott2001studying}. 
The coefficient $\Dcm$ can be measured by non-single-molecule, but super-resolution fluorescence microscopy \cite{gahlmann2014exploring}. The coefficient $\DR$ can be measured by following the expanding radius of a fluorescent domain. We address all four coefficients below. In particular, the ratio $\Dcm/\DR$ serves as a characteristic of the extent to which the assembly moves collectively before it disperses, i.e., its persistence as a compact object.

One should mention also the gradient diffusion coefficient, $\Dgrad$, which is the one entering the diffusion equation for the continuous concentration of particles (as measured for example by fluorescence recovery after photobleaching (FRAP) \cite{lippincott2001studying}). This coefficient is not equivalent to $\DR$ because it is measured in the laboratory reference frame and thus is affected also by the CM motion.

When the motions of particles are decoupled, $\Ds =\DR =\Dgrad =D_0$, and $\Dcm=D_0/N$, where $N$ is the number of particles. Hence, for assemblies containing more than a dozen particles, $\Dcm$ is negligible in comparison to the other coefficients. As a result, the assembly disperses before its CM had the time to move appreciably. As we shall see below, the correlations in fluid membranes lead to strikingly different behavior\,---\,$\Dcm$ is similar or even larger than $\DR$, giving the assembly the opportunity to diffuse collectively before it spreads out. It was recognized before that membrane-embedded extended objects, such as polymer chains, have a $\Dcm$ comparable to the diffusion coefficient of their constituent monomers 
\cite{muthukumar1985brownian,komura1995diffusion,ramachandran2011dynamics,seki2011diffusion}. In the present work we focus on the implications of the underlying strong correlations for the dynamics of a dispersing assembly of inclusions. Unlike those earlier works, the dispersing assembly is inherently out of equilibrium and its statistical distributions are time-dependent.

\section*{Methods}

\subsection*{Model}

\begin{figure}[hbt!]
\centering
\includegraphics[width=0.45\linewidth]{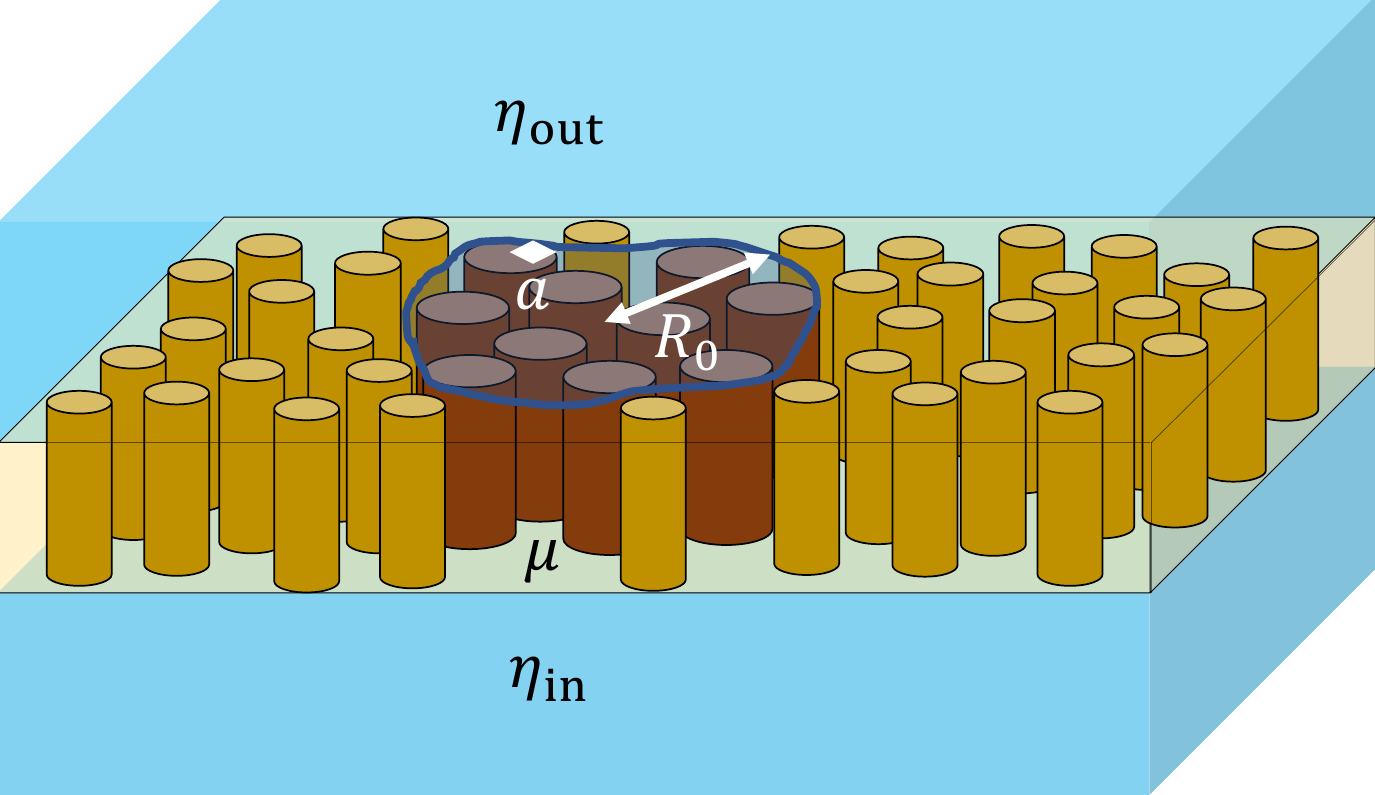}
\caption{Illustration of the model and its parameters.}
\label{fig:system}
\end{figure}

Following the SD model, we treat the membrane as a flat fluid layer of two-dimensional (2D) viscosity $\mu$, in contact on its two sides with two fluids of three-dimensional (3D) viscosities $\eta_{\rm in}$ and $\eta_{\rm out}$. See Fig.~\ref{fig:system}. The SD length is defined as
\begin{equation}
    \kappa^{-1} = \mu/(\eta_{\rm in}+\eta_{\rm out}).
\end{equation}
The diffusion coefficient of an isolated disk-like inclusion of radius $a\ll\kappa^{-1}$ is given by SD as
\begin{equation}
    D_0 = \frac{\kT}{4\pi\mu} \left( \ln \frac{2}{\kappa a} - \gamma \right),\label{SD D0}
\end{equation}
where $\kT$ is the thermal energy, and $\gamma\simeq 0.58$ is Euler's constant.

We consider an assembly of $N$ such inclusions, distributed initially over an area of typical radius $R_0\ll\kappa^{-1}$, as illustrated in Fig.~\ref{fig:system}. (The particular distribution and exact definition of $R_0$ will be presented below.) For simplicity, we assume an isotropic distribution.
The positions of the particles are $\{\vecr^i\}$, $i=1,\ldots,N$. We use Latin indices to represent the particles, and Greek indices to indicate the in-plane coordinates $(x,y)$. 

When forces $\{\vecF^i\}$ are exerted on the particles, their
velocities respond linearly according to 
\begin{equation}
    v_{\alpha}^{i} = \indB_{\alpha\beta}^{ij}\left(\left\{ \vecr^{k=1...N}\right\} \right) F_{\beta}^{j},
\label{vBF}
\end{equation}
written in the convention of summation over repeated indices. The many-particle mobility tensor $\tenB$, which depends on the particles' instantaneous configuration, characterizes the self-response of a particle to the force acting on it ($i=j$), as well as its response to the forces acting on the others ($i\neq j$). The latter arises from the flow-induced coupling between the particles. We use the following tensor:
\begin{gather}
    \indB_{\alpha\beta}^{i=j} = B_{0}\delta_{\alpha\beta}=\frac{1}{4\pi \mu}\left(\ln \frac{2}{\kappa a}-\gamma\right) \delta_{\alpha\beta},
    \nonumber\\
    \indB_{\alpha\beta}^{i\neq j}=\frac{1}{4\pi \mu} \left[ \left( \ln \frac{2}{\kappa r^{ij}} - \gamma-\frac{1}{2}\left(1-\frac{2a^{2}}{\left(r^{ij}\right)^{2}}\right)\right)\delta_{\alpha\beta} + \left(1-\frac{2a^{2}}{\left(r^{ij}\right)^{2}}\right)\frac{r_{\alpha}^{ij} r_{\beta}^{ij}}{\left(r^{ij}\right)^{2}}\right],
    \label{Btensor}
\end{gather}
where $\vecr^{ij}\equiv\vecr^j-\vecr^i$.
This tensor, derived in Ref.~\cite{sokolov2018many}, is analogous to the Rotne-Prager-Yamakawa tensor for 3D suspensions \cite{rotne1969variational,yamakawa1970transport}. 
Although it contains the leading two orders in large inter-particle separation, it is guaranteed to be positive-definite for any configuration where all $r^{ij}>2a$. In addition, the tensor is valid in the limit $r^{ij}\ll \kappa^{-1}$. Under these approximations the self components are configuration-independent and given simply by the SD mobility, Eq.~\ref{SD D0}. The coupling components contain the well-known, leading logarithmic terms \cite{bussell1992resistivity,oppenheimer2009correlated} mentioned in the introduction, as well as the next-order (quadratic) correction in small $a/r^{ij}$ \cite{sokolov2018many}.

When random thermal forces act on the particles, they perform Brownian motions, which gradually disperse the assembly. These random velocities are correlated according to Eqs.~\ref{vBF} and \ref{Btensor}, leading to configuration-dependent complex dynamics.

\subsection*{Analytical calculations}

The CM position is given by $\Rcm=N^{-1}\sum_{i=1}^N\vecr^i$. We denote by $\vDelta^{i}\left(t\right)$ the displacement of particle $i$ from an initial position $\vecr_{0}^{i}$ to its position after time $t$, $\vecr^{i}$. The corresponding CM displacement is $\vDelta_{\rm CM}\left(t\right)=N^{-1}\sum_i \vDelta^i\left(t\right)$.

The CM diffusion coefficient is related to its mean square displacement (MSD),
\begin{eqnarray}
    4\int_0^t\Dcm\left(s\right)ds\equiv \left\langle \vDelta_{\rm CM}\cdot\vDelta_{\rm CM}\right\rangle = N^{-2} \sum_{i=1}^N \left\langle \vDelta^{i}\cdot\vDelta^{i}\right\rangle + 2 N^{-2} \sum_{i=1}^{N}\sum_{j>i}^{N}\left\langle\vDelta^{i}\cdot\vDelta^{j}\right\rangle.\label{4Dcmt}
\end{eqnarray}
The first term on the right-hand side relates to the self-diffusion of the individual particles, whereas the second one arises from the coupled diffusion. These displacement correlations are determined by certain diffusion coefficients, $\left\langle\vDelta^{i}\cdot\vDelta^{j}\right\rangle=2\int_0^t\left(\indD_{xx}^{ij}+\indD_{yy}^{ij}\right)dt$. In fact, this diffusion tensor is directly related to the mobility tensor of Eq.~\ref{Btensor} through the Einstein-Smoluchowski relation, $\indD_{\alpha\beta}^{ij} = \kT \indB_{\alpha\beta}^{ij}$.
Using the tensor components of Eq.~\ref{Btensor} we identify the CM diffusion coefficient. Without correlations it is given by a constant
\begin{equation}
    D_{\rm CM}^{\rm nc} = N^{-1}D_{0}=N^{-1}\frac{k_{\rm B}T}{4\pi\mu} \left(\ln\frac{2}{\kappa a}-\gamma\right),\label{Dcm no corr}
\end{equation}
and with the flow-mediated couplings, by
\begin{equation}
    \Dcm\left(t\right)=\const\left(\frac{1}{N}\ln\frac{2}{\kappa a}-\gamma +\frac{2}{N^{2}}\sum_{i=1}^{N}\sum_{j>i}^{N}\ln\frac{2}{\kappa r^{ij}\left(t\right)}\right).\label{Dcm with sums}
\end{equation}
Note that the correction terms $\sim a^2/\left(r^{ij}\right)^{2}$ appearing in Eq.~\ref{Btensor} cancel out in $\Dcm$. It is important to note also that the time-dependence of $\Dcm(t)$ does not arise from memory, as Eq.~\ref{Btensor} describes a strictly instantaneous response. The coefficient depends on the instantaneous configuration, which changes with time.

The squared radius of gyration is defined as $\RG^2=N^{-1} \sum_{i=1}^{N}\left(\vecr^i-\Rcm\right)^2$. Thus,
\begin{eqnarray}
    \left\langle \RG^2\right\rangle&\equiv& \frac{1}{N}\left\langle\sum_{i=1}^{N}\left(\vecr^i-\Rcm\right)^2\right\rangle=
    R_0^2+\frac{1}{N}\sum_{i=1}^{N}\left\langle \vDelta^{i}\cdot\vDelta^{i}\right\rangle -2\left\langle \frac{1}{N}\sum_{i=1}^{N}\vDelta^{i}\cdot\frac{1}{N}\sum_{j=1}^{N}\vDelta^{j}\right\rangle +\left\langle \vDelta_{\rm CM}\cdot\vDelta_{\rm CM}\right\rangle  \nonumber\\&=&R_0^2+4D_0t-2\left\langle \vDelta_{\rm CM}\cdot\vDelta_{\rm CM}\right\rangle +4\int_0^t\Dcm\left(s\right)ds=R_0^2+4D_0t-4\int_0^t\Dcm\left(s\right)ds\label{Rg std dev},
\end{eqnarray}
where $R_0\equiv \RG\left(t=0\right)$. From this equation we identify the variance of the gyration radius, $\left\langle \RG^2\right\rangle - R_0^2\equiv 4\int_0^t\DR\left(s\right)ds$, and the diffusion coefficient of the radius of gyration, 
\begin{equation}
    \DR =D_0-\Dcm. \label{DRg as Dcm}
\end{equation}
Usually $\DR \simeq D_0\gg\Dcm$. This is not the case in the present work, where $\Dcm$ is comparable to $D_0$, and therefore also to $\DR$. Furthermore, if $\Dcm>D_0/2$, the cluster disperses more slowly than it moves collectively, $\Dcm>\DR$. Like $\Dcm$, $\DR$ is also configuration-dependent (through $\Dcm$). Substituting $\Rcm=N^{-1}\sum_{i=1}^N\vecr^i$ in Eq.~\ref{Rg std dev}, we similarly obtain the variance of the inter-particle separation $\vecr^{ij}$ in terms of $\left\langle \RG^2\right\rangle$, which is
\begin{equation}
    \left\langle \vecr^{ij}\cdot\vecr^{ij}\right\rangle=2R_0^2+8D_0t-8\int_0^t\Dcm\left(s\right)ds=2R_0^2+8\int_0^t\DR\left(s\right)ds.\label{rij std dev}
\end{equation}

The other diffusion coefficients mentioned in the Introduction are less relevant to our main effect, as they are all comparable to $D_0$. Under the assumption of the pairwise-additive hydrodynamic interactions (see Eq.~\ref{Btensor}) and the lack of direct interaction, the self-diffusion coefficient $\Ds$ is equal to $D_0$. The two might differ due to direct and many-body interactions, as captured for example by changes in the effective viscosity of the heterogeneous fluid~\cite{camley2014fluctuating}. In fact, the more general and accurate version of Eq.~\ref{DRg as Dcm} is $\DR =\Ds -\Dcm$. In addition, as will be shown below, the entire contribution of the self-terms to $\Dcm$ is negligible during the relevant stage of dispersal (see Eq.~\ref{Dcm final}). The gradient diffusion coefficient $\Dgrad$, as well, will generally deviate from $D_0$ and $\Ds$ due to correlations that go beyond our present model (for more details, see, for example, Ref.~\cite{pusey89}).

\subsection*{Simulation method}

We use the technique of Brownian dynamics with hydrodynamic interactions \cite{ermak1978brownian,brady1988stokesian}. The simulation takes place in a square box of side $L\equiv 1$, with periodic boundary conditions. Initially, a set of $N$ disks of radius $a=0.005$ are placed randomly, but without overlap, within a circular area of radius $R_0=0.1$; see Fig.~\ref{fig:snapshots} for examples of initial configurations. At each  step a set of $N$ stochastic but correlated forces (see below) are applied to the particles. The particles' velocities are then calculated using the many-particle mobility tensor of Eq.~\ref{Btensor}, with $\kappa^{-1}=5$ and $\mu\equiv 1$. The particles' displacements to the next configuration are then calculated using the simplest Ito convention. A displacement that makes a disk overlap another is rejected, and that particle is not moved. The positive-definite tensor (Eq.~\ref{Btensor}) guarantees dynamic stability \cite{sokolov2018many}. 

Each run consists of 1000 steps, and the results are averaged over 100 runs (with different initial configurations). These results include the MSD of individual particles, the MSD of the center of mass, and the difference in the squared radius of gyration from its initial value. The corresponding diffusion coefficients, $\Ds$, $\Dcm$, and $\DR$, are obtained from the slopes of these mean-squared displacements as functions of $t$. Errors are estimated by comparing the linear fit using the first 500 steps and all 1000 steps. We simulated different numbers $N$ while keeping $R_0$ fixed, thus sampling different initial densities.

The main difficulty lies in producing the correlated random forces $\vecf^i$, $i=1,\ldots N$, such that the fluctuation-dissipation theorem is obeyed, $\langle f^i_\alpha(t) f^j_\beta(t') \rangle = 2 \kT (\indB^{-1})^{ij}_{\alpha\beta} \delta(t-t')$ \cite{ermak1978brownian,brady1988stokesian}. Although the procedure is known, we repeat it here to help future implementations. At each step we perform the following: (1)  calculate the $2N\times 2N$ mobility matrix $\tenB$ using Eq.~\ref{Btensor}; (2) diagonalize $\tenB$ into $\hat\indB_{pq}=b_p\delta_{pq}$, $p,q=1,\ldots,2N$; (3) construct the diagonal matrix $\hat\indA_{pq}=a_p\delta_{pq}$ where $a_p=\sqrt{2\kT/(b_p dt)}$ and $dt$ is the time interval associated with a simulation step; (4) transform $\hat\tenA$ to the basis of the original $\tenB$, obtaining the non-diagonal matrix $\tenA$; (5) produce a set of $2N$ independent standard Gaussian random variables, $f_{0,p}$, such that $\langle f_{0,p}\rangle = 0$ and $\langle f_{0,p} f_{0,q}\rangle = \delta_{pq}$ ; (6) obtain the correlated forces by applying $\tenA$ to the uncorrelated forces, $f_p=\indA_{pq}f_{0,q}$.

The results are translated into physical units using $\mu=10^{-6}$~poise$\cdot$cm \cite{brown2011continuum}, $T=300$~K, and $\kappa^{-1}=500$~nm. The latter implies that the simulation box side is equivalent to $100$~nm, the initial domain radius to $10$~nm, and the disk diameter to $1$~nm.

Movies of typical simulations without and with flow-mediated interactions are found in the Supplementary Material.

\section*{Results}

\subsection*{Analytical results}

Our main focus is the CM diffusion coefficient, $\Dcm$. The expression for that diffusion coefficient given a certain configuration is found in Eq.~\ref{Dcm with sums}. We look for the ensemble average over all configurations, $\left\langle\Dcm\right\rangle$. (For brevity, we will continue to use the symbol $\Dcm$, omitting the angular brackets.) In Eq.~\ref{Dcm with sums} we identify $\ln\left(\kappa r^{ij}\right)$ as the only term that should be averaged. To perform the averaging, we are required to find the statistical distribution of inter-particle separations, $\vecr^{ij}$. Unlike earlier works on membrane-embedded polymers, this statistical distribution changes over time as the assembly progressively disperses. As a result, $\left\langle\ln\left(\kappa r^{ij}\right)\right\rangle$ itself depends on $\Dcm$ (see Eq.~\ref{rij std dev}). This will lead to a self-consistent equation for $\Dcm\left(t\right)$.

The inter-particle separation distribution, $G\left(\vecr^{ij},t\right)$, is to a good approximation a Gaussian with a variance given by Eq.~\ref{rij std dev}. Applying this distribution to the required average, we obtain from Eq.~\ref{Dcm with sums} 
\begin{gather}
    \Dcm\left(t\right)=\const\left(\frac{1}{N}\ln\frac{2}{\kappa a} +\frac{N-1}{N}\left(\ln\frac{2}{\kappa R\left(t\right)}-b\right)-\gamma\right),\nonumber\\
    b\equiv \int d^{2}\vrho G\left(\vrho \right) \ln \rho = \frac{1}{2} \left(\ln 2 - \gamma\right)\simeq 0.058,\label{Dcm not approx}
\end{gather}
where $R\left(t\right)\equiv\sqrt{\left\langle \RG^2\right\rangle}$, such that $R\left(t=0\right)=R_0$, and $\vrho\equiv\vecr^{ij}/R$. Using a different (non-Gaussian) distribution or different domain shape will change the geometric factor $b$ into a time-dependent function $b\left(t\right)$ (see Supplementary Material). The time dependence is weak as the distribution tends to a Gaussian eventually. Since $N\gg 1$, we can rearrange Eq.~\ref{Dcm not approx} into
\begin{gather}
    \Dcm\left(t\right)=\const\left(\ln\frac{2}{\kappa \bar{R}\left(t\right)} -\gamma\right),\label{Dcm final}
\end{gather}
where $\bar{R}=e^{b}R\simeq 1.06R$. Comparing Eq.~\ref{Dcm final} with $D_0$ of Eq.~\ref{SD D0}, we see that the assembly's CM diffuses effectively like a rigid disk with effective radius $\bar{R}$. This result is similar to the diffusion coefficient of a circular liquid domain found in Ref.~\cite{seki2011diffusion} (see also Supplementary Material). The main difference is that our effective size changes over time. The diffusion coefficient of gyration could be written in terms of $\bar{R}\left(t\right)$,
\begin{equation}
    \DR \left(t\right)=\const\ln\frac{\bar{R}\left(t\right)}{a},\label{DR final}
\end{equation}
which, surprisingly, is independent of the SD length.

Substituting Eq.~\ref{Rg std dev} in $\bar{R}=e^{b}\sqrt{\left\langle \RG^2\right\rangle}$ gives the self-consistent equation for $\Dcm\left(t\right)$,
\begin{equation}
    \Dcm\left(t\right)=\frac{\kT}{8\pi\mu}\left(-\ln\left(\frac{\kappa^{2}}{2}\left(R_{0}^{2}+4D_{0}t-4\int_0^{t}\Dcm\left(s\right)ds\right)\right)-\gamma\right).\label{Dcm sce}
\end{equation}
Because of the model's assumptions, this equation is only valid when the inter-particle distances are smaller than the SD length, $R\ll\kappa^{-1}$. Since $R^2=R_{0}^{2}+4D_{0}t-4\int_0^{t}\Dcm\left(s\right)ds>4\DR\left(0\right)t$, the time during which the model is valid is $t\ll\frac{\kappa^{-2}}{\DR\left(0\right)}$. Equation~\ref{Dcm sce} has the exact solution,
\begin{gather}
    \Dcm\left( t\right)=D_0-\frac{\kT}{8\pi\mu}\text{Ei}^{-1}\left(\frac{R_{\rm eff}^2 +4D_{\rm exp}t}{a^2}\right),\label{exact}\\
    R_{\rm eff}^{2}\equiv a^{2}\text{Ei}\left(\ln\frac{2R_{0}^2}{a^2}-\gamma\right),\ \ 
    D_{\rm exp}\equiv\frac{\kT e^{-\gamma}}{4\pi\mu}\nonumber,
\end{gather}
where $\text{Ei}\left(x\right)=\int_{-\infty}^{x}\frac{e^{\zeta}}{\zeta}d\zeta$ is the exponential integral function, and $\text{Ei}^{-1}\left(x\right)$ is its inverse.  Notice that $\Dcm$'s time dependence is not a function of $R_{0}^{2}+4D_{0}t$ as might have been expected, but rather has another initial effective square radius $R_{\rm eff}^{2}<R_{0}^{2}$, and an effective expansion diffusion coefficient $D_{\rm exp}<D_{0}$ (the inequalities arise from the assumption that $\kappa^{-1}\gg R_{0}\gg a)$. Curiously, $D_{\rm exp}$ depends only on $T$ and $\mu$, and is independent of any length scale ($\kappa$, $a$, $R_0$).

\begin{figure}[hbt!]
\centering
\includegraphics[width=0.45\linewidth]{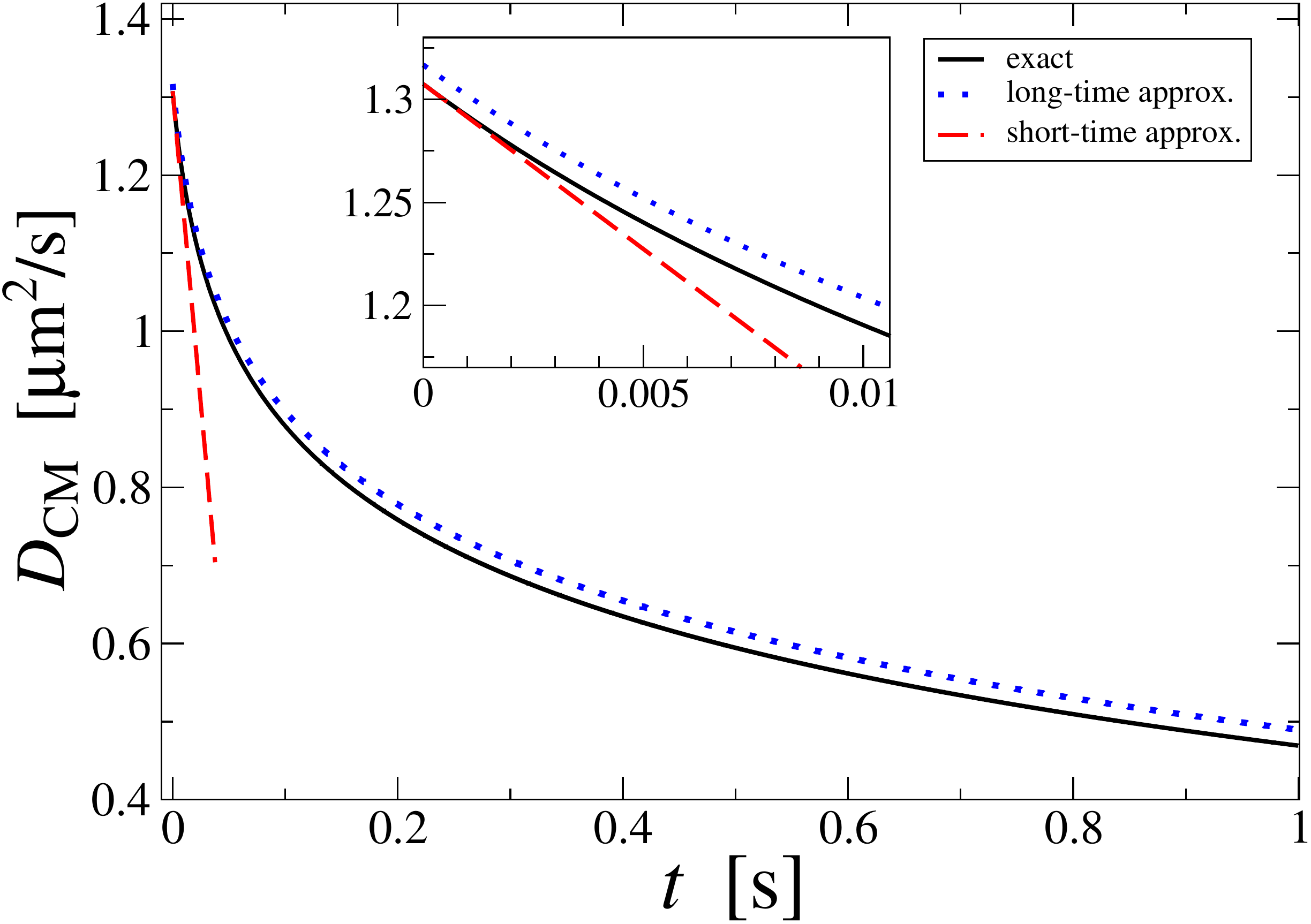}
\caption{Center-of-mass diffusion coefficient vs.\ time, as obtained analytically for a Gaussian-distributed assembly of particles. The solid line shows the exact solution (Eq.~\ref{exact}), the dashed line is the short-time asymptote (Eq.~\ref{short}), and the dotted line is the long-time approximation (Eq.~\ref{long}). The inset focuses on the experimentally relevant short-time region. Parameter values: $a=0.5$~nm, $R_0=10$~nm, $\kappa^{-1}=500$~nm, $\mu=10^{-6}$~poise$\cdot$cm, $T=300$~K.}
\label{fig:analytic}
\end{figure}

This solution is presented in Fig.~\ref{fig:analytic} (solid line). The decrease of $\Dcm$ with time implies that the CM performs sub-diffusion. This could be verified directly in Eq.~\ref{exact}, since $\text{Ei}^{-1}\left(x\right)$ is a monotonically increasing function. The physical origin of the sub-diffusion lies in the weakening of correlations as the assembly disperses. The time dependence is only logarithmic (see Eq.~\ref{long} below), and its experimental (and numerical) relevance is therefore limited.

To clarify Eq.~\ref{exact} further, we present two asymptotic limits of the exact solution. In the short-time regime, we assume a perturbed initial CM diffusion coefficient, $\Dcm\left(t\right)= \Dcm\left(0\right)+\varepsilon_{\rm CM}\left(t\right)$. To leading order in small $\varepsilon_{\rm CM}\left(t\right)/\DR\left(0\right)$, we get  
\begin{equation}
    \Dcm\left(t\right)\simeq\Dcm\left(0\right)-\frac{\kT}{4\pi\mu R_0^2}\DR\left(0\right)t. \label{short}
\end{equation}
This linear expression is valid for  $t\ll 4\pi\mu R_{0}^{2}/\kT$. Since this value is only one order of magnitude bigger than $R_0^2/D_0$, at longer times the assembly has already dispersed significantly. Thus, $\Dcm(0)$ and its leading correction given by Eq.~\ref{short} provide a good approximation over the relevant time scales. This will be demonstrated also in the following section. It is noteworthy that the exact $\Dcm$ is always larger than the short-time approximation. (See Fig.~\ref{fig:analytic}, dashed line.)

The second, "long-time" asymptote uses the expansion $\text{Ei}^{-1}\left(x\right)\simeq\ln\left(x\ln x\right)$ for large $x$. We put the term "long-time" in quotation marks because, on the one hand, the corrections to the expansion at any order are never small due to their logarithmic nature \cite{pecina1986function}. On the other hand, since we always have $R_{\rm eff}\gg a$, the argument in the expansion, indeed, is always large. Hence, the approximation
\begin{equation}
    \Dcm\left(t\right)\simeq\const\left(\ln\left(\frac{2}{\kappa\sqrt{R_{\rm eff}^2+4D_{\rm exp}t}}\right)-\gamma -\frac{1}{2}\ln\left(\ln\left(\frac{R_{\rm eff}^2+4D_{\rm exp}t}{a^2}\right)\right)\right) \label{long}
\end{equation}
has a wide range of validity, but at the same time is never very accurate. Figure~\ref{fig:analytic} demonstrates this unusual situation.


Substituting Eqs.~\ref{exact}--\ref{long} in Eq.~\ref{DRg as Dcm}, we readily get the corresponding expressions for the gyration diffusion coefficient,
\begin{gather}
    \DR\left(t\right)=\frac{\kT}{8\pi\mu}\text{Ei}^{-1}\left(\frac{R_{\rm eff}^2 +4D_{\rm exp}t}{a^2}\right),\label{DR exact}\\
    \DR\left(t\right)\simeq\DR\left(0\right)+\frac{\kT}{4\pi\mu R_0^2}\DR\left(0\right)t,\ \ \text{ short-time approximation}\label{DR short}\\
    \DR\left(t\right)\simeq\const\left(\ln\left(\frac{\sqrt{R_{\rm eff}^2+4D_{\rm exp}t}}{a}\right)+\frac{1}{2}\ln\left(\ln\left(\frac{R_{\rm eff}^2+4D_{\rm exp}t}{a^2}\right)\right)\right),\ \ \text{ "long-time" approximation.}
\label{DR long}\end{gather}
Note again that $\DR$ is independent of $\kappa$.

Equations~\ref{exact}--\ref{long} are our central results. Comparing to Eq.~\ref{Dcm no corr} we can appreciate the remarkable difference that arises from flow-induced correlations. The strong $N^{-1}$ suppression of $\Dcm$ is replaced by a much weaker logarithmic dependence on the cluster's radius. When the particles become indefinitely far apart, Eq.~\ref{Dcm no corr} will eventually apply. This will occur when the inter-particle distance becomes comparable to $\kappa ^{-1}$ (after time of order $\kappa^{-2}/\left(\DR\left(0\right)\right)$). This limit lies beyond the assumptions of the model, and is not captured by Eq.~\ref{exact}. It is not of much interest, though, as it implies that the cluster has already disintegrated.

The self-consistent calculation presented above can be repeated for an assembly of spherical particles in 3D. The effect is much weaker. Since the hydrodynamic interactions in this case make $\Dcm$ inversely proportional to $R$ (as in the Zimm dynamics of polymers~\cite{rubinstein2003polymer}), it becomes comparable to $\DR$ only for significantly smaller cluster radii (i.e., denser clusters).








\subsection*{Simulation results}

\begin{figure}[hbt!]
\begin{center}
\includegraphics[width=0.42\linewidth]{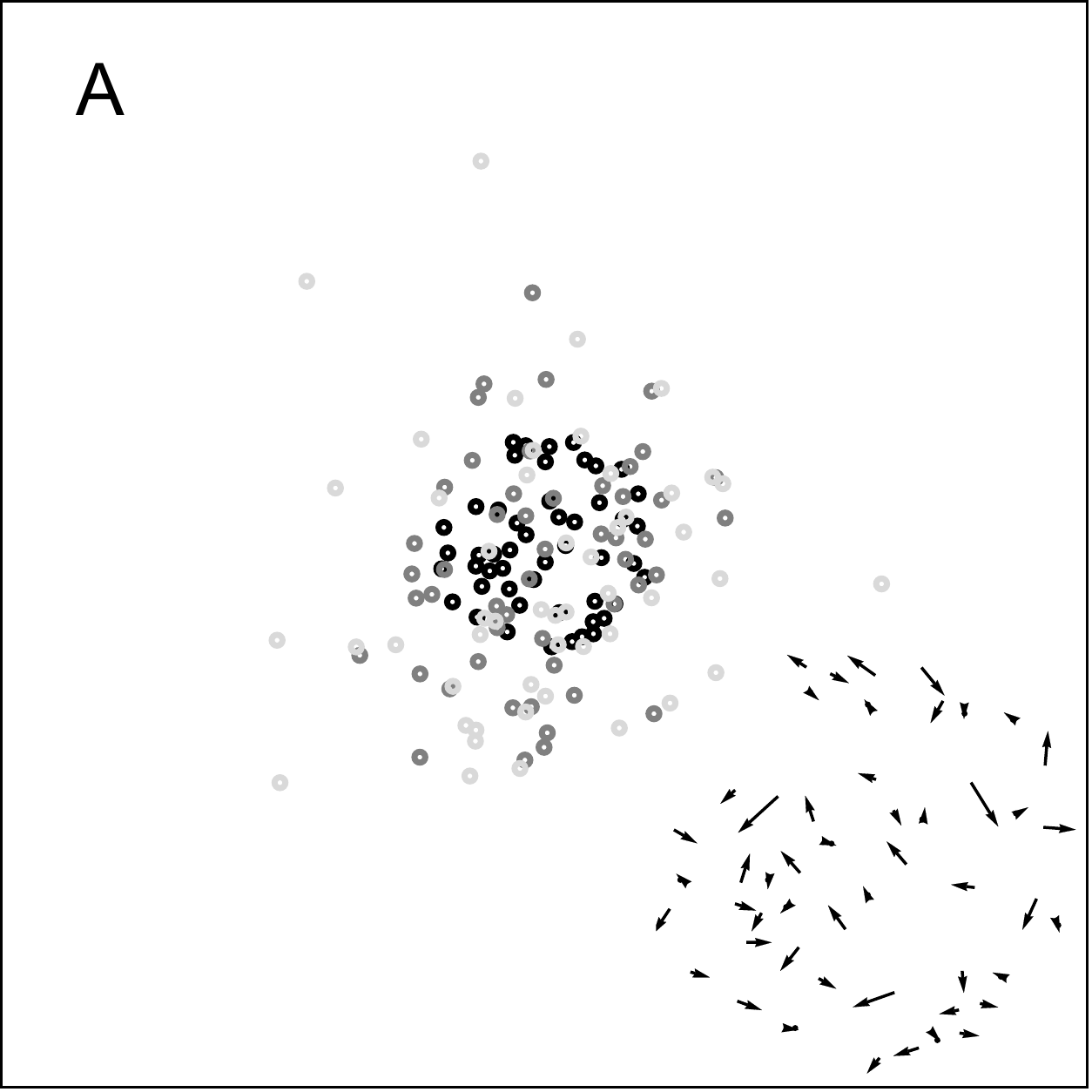}
\hspace{0.4cm}
\includegraphics[width=0.42\linewidth]{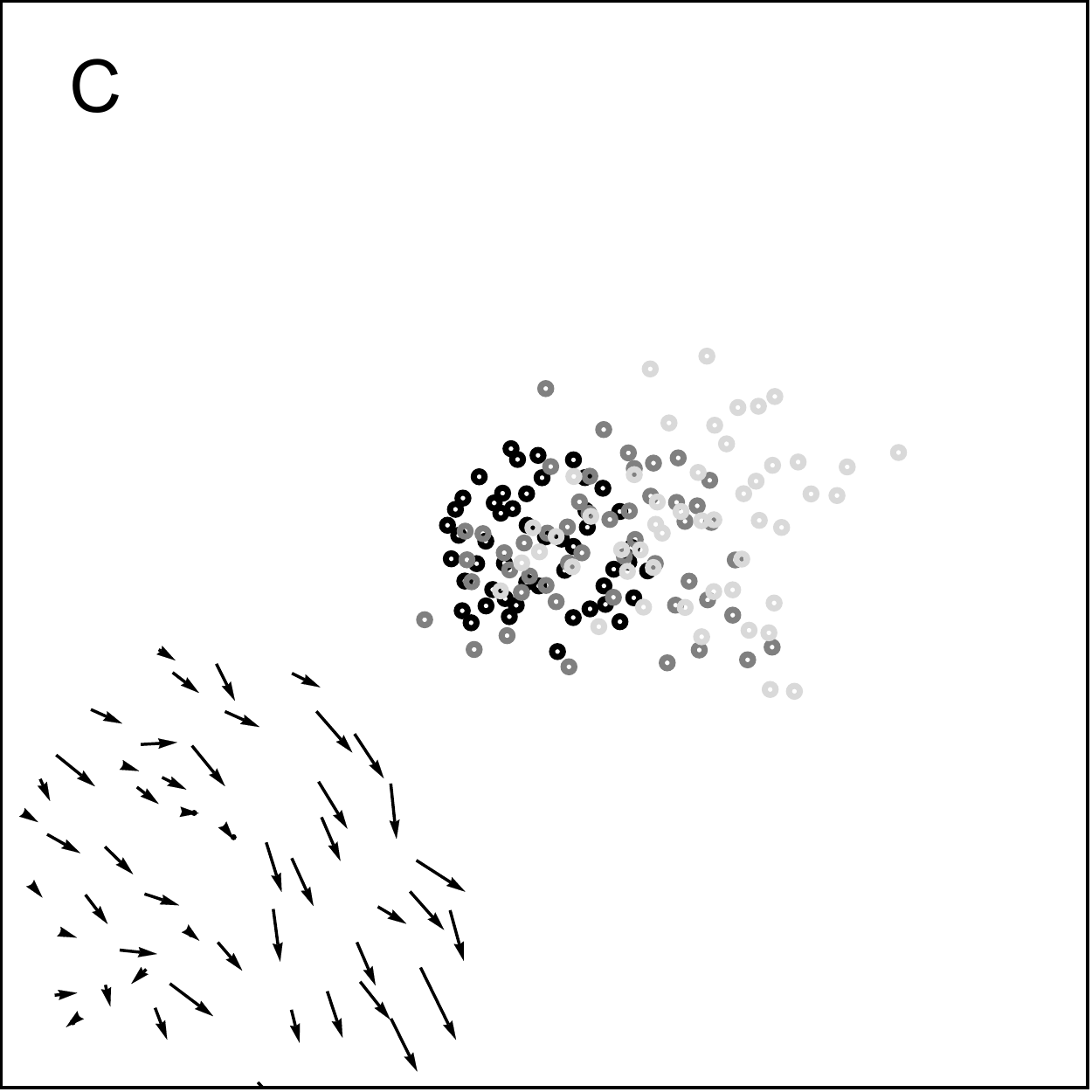}
\end{center}
\vspace{0.5cm}
\begin{center}
\includegraphics[width=0.45\linewidth]{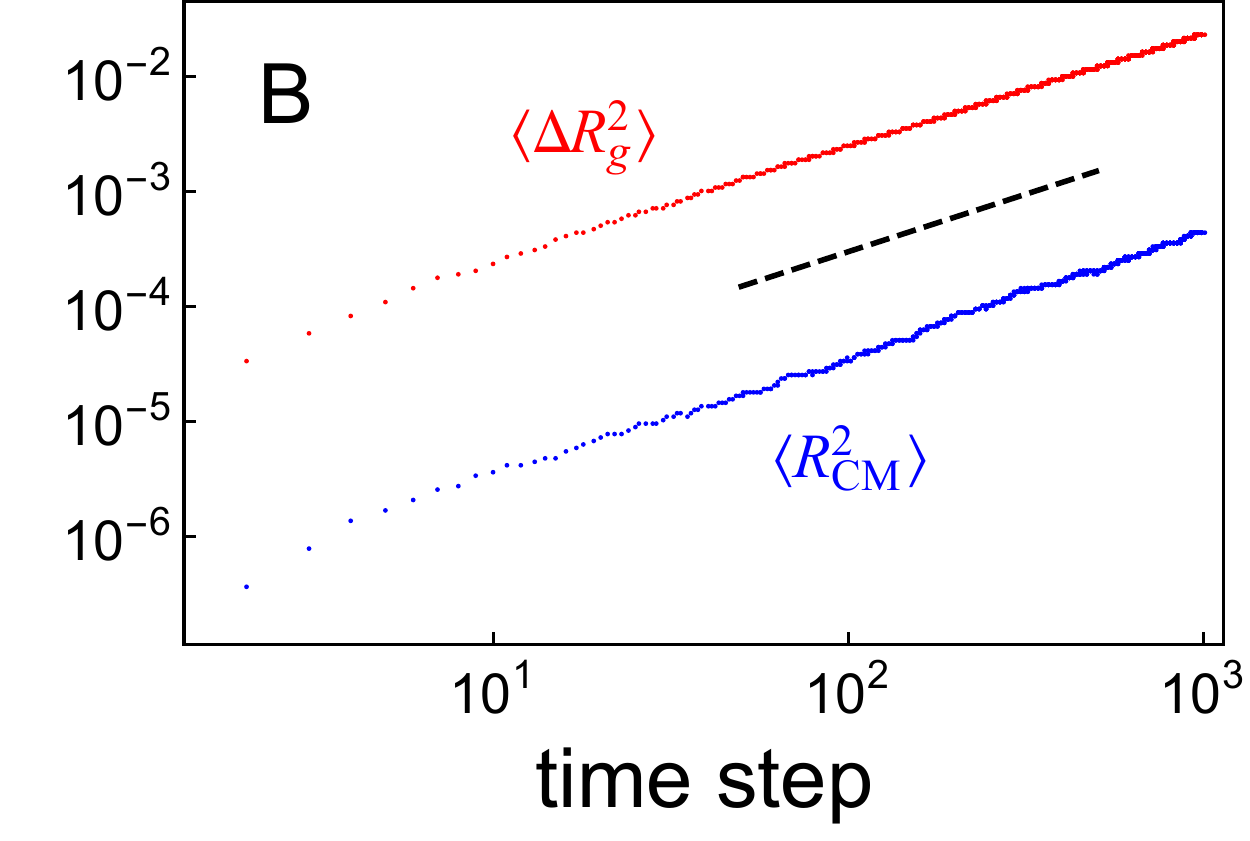}
\hspace{0.2cm}
\includegraphics[width=0.45\linewidth]{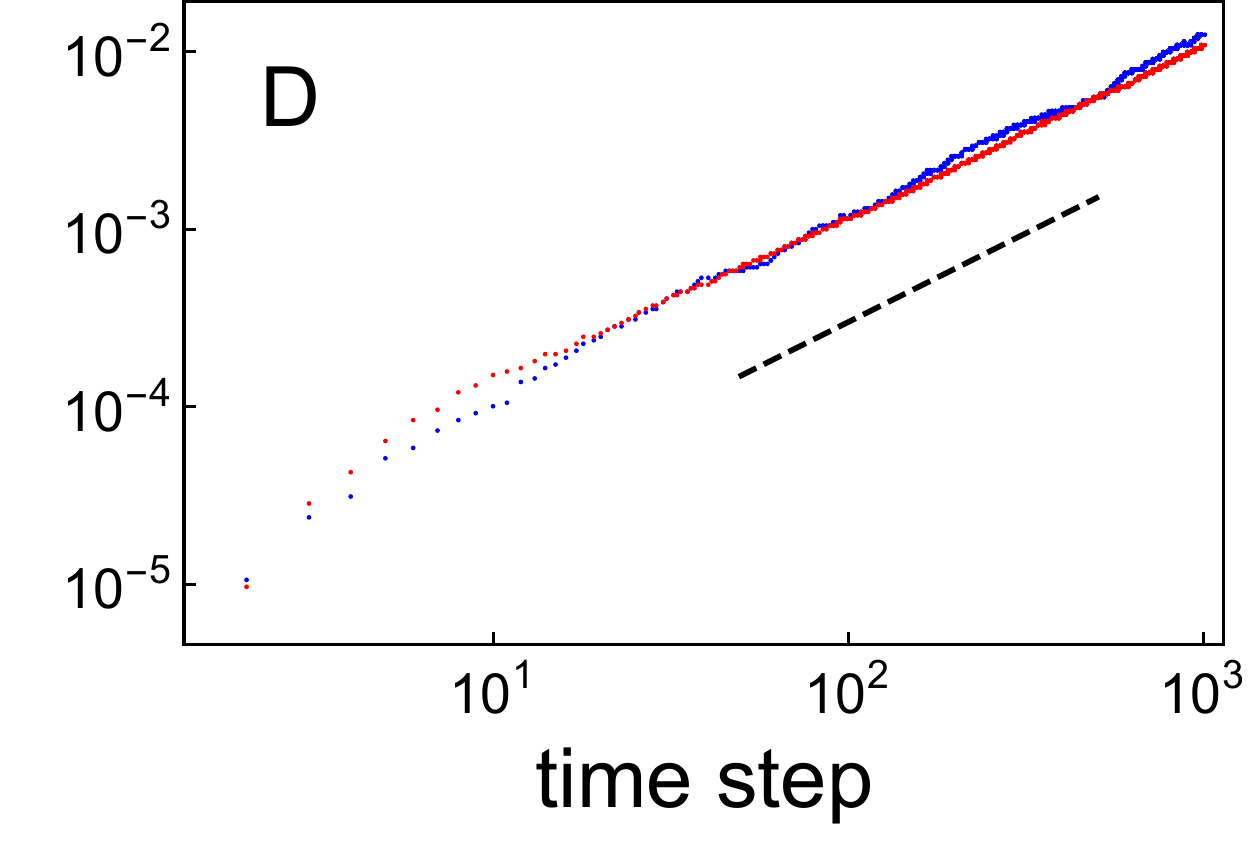}
\end{center}
\caption{Simulated time evolution of assemblies made of 50 particles, without (A,B) and with (C,D) flow-mediated correlations. Panels A and C show snapshots of the initial configuration (black) and the configurations after 500 (gray) and 1000 (light gray) time steps. Arrows at the corners show particle displacements from the initial configuration to the one after 10 time steps (the scale is larger; the patch of arrows is the same size as the black patch of particles). Panels B and D show on a log-log scale the corresponding mean-square displacement of the center of mass (lower, blue) and change in mean-square radius of gyration (upper, red), averaged over 100 runs. The unit of length is the simulation box size. The dashed line has a slope of 1 for comparison.}
\label{fig:snapshots}
\end{figure}

Figure \ref{fig:snapshots} demonstrates the main effect that we wish to highlight. Panels~A and C show snapshots from simulations of two 50-particle assemblies, without and with the flow-induced correlations. (See also the movies in the Supplementary Material.) While the uncorrelated assembly disperses before it had the time to move much as a whole (Panel A), the correlated one shows a vivid displacement of its center of mass (Panel C). These panels present also the particles' displacement vectors from the initial positions to the positions after 10 time steps, demonstrating the directionality induced by the correlations. Panels~B and D show the CM MSD and the difference in the mean-square radius of gyration from its initial value, as a function of time, averaged over 100 runs. In the case of the correlated assembly the CM MSD is much larger and becomes comparable to the mean-square radius of gyration. No consistent change of slope could be discerned in the CM MSD curves. Thus, the time-dependent logarithmic correction to the initial CM diffusion coefficient, obtained analytically above, is too small to be resolved in the present simulations. In the following analysis, therefore, we will consider the CM diffusion coefficient as being constant.

We now proceed to a more detailed analysis of the dynamics. All comparisons to theoretical predictions are without any fitting parameters.

\begin{figure}[hbt!]
\centering
\includegraphics[width=0.45\linewidth]{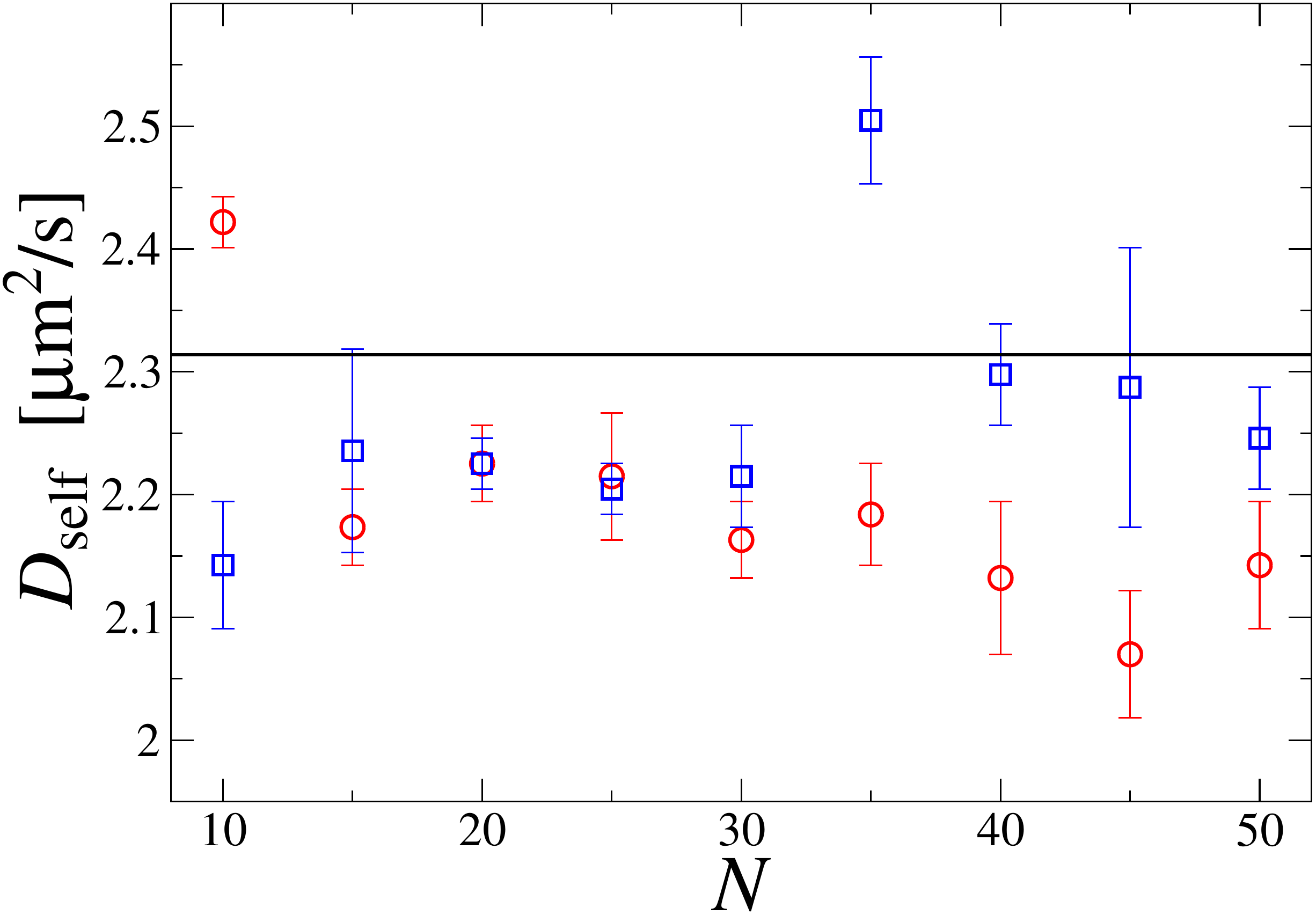}
\caption{Self-diffusion coefficient as a function of particle number, obtained from simulations of  uncorrelated (red circles) and correlated (blue squares) assemblies. Parameter values are as in Fig.~\ref{fig:analytic}. The horizontal line shows the value of the theoretical SD diffusion coefficient of an isolated particle (Eq.~\ref{SD D0}) for these parameters, $D_0=2.31$~$\mu$m$^2$/s.}
\label{fig:Dself}
\end{figure}

Figure \ref{fig:Dself} shows the simulation results for the self-diffusion coefficient of individual particles within the assemblies. The $\Ds$ values are scattered within a 10\% difference around the theoretical value of $D_0$ (solid line), without a consistent dependence on assembly size or flow-mediated correlations. This is in line with the prediction of $\Ds=D_0$ for our model (Eq.~\ref{Btensor}). The prediction relates to the short-time self-diffusion coefficient in the absence of interactions \cite{pusey89}, whereas the simulated particles have excluded-volume interactions and their MSD's are measured over the entire simulation time. The deviations found in Fig.~\ref{fig:Dself}, therefore, are to be expected, and the overall trend of slightly smaller coefficients in the simulations probably arises from the exclusion (crowding)  effect.

\begin{figure}[hbt!]
\centering
\includegraphics[width=0.45\linewidth]{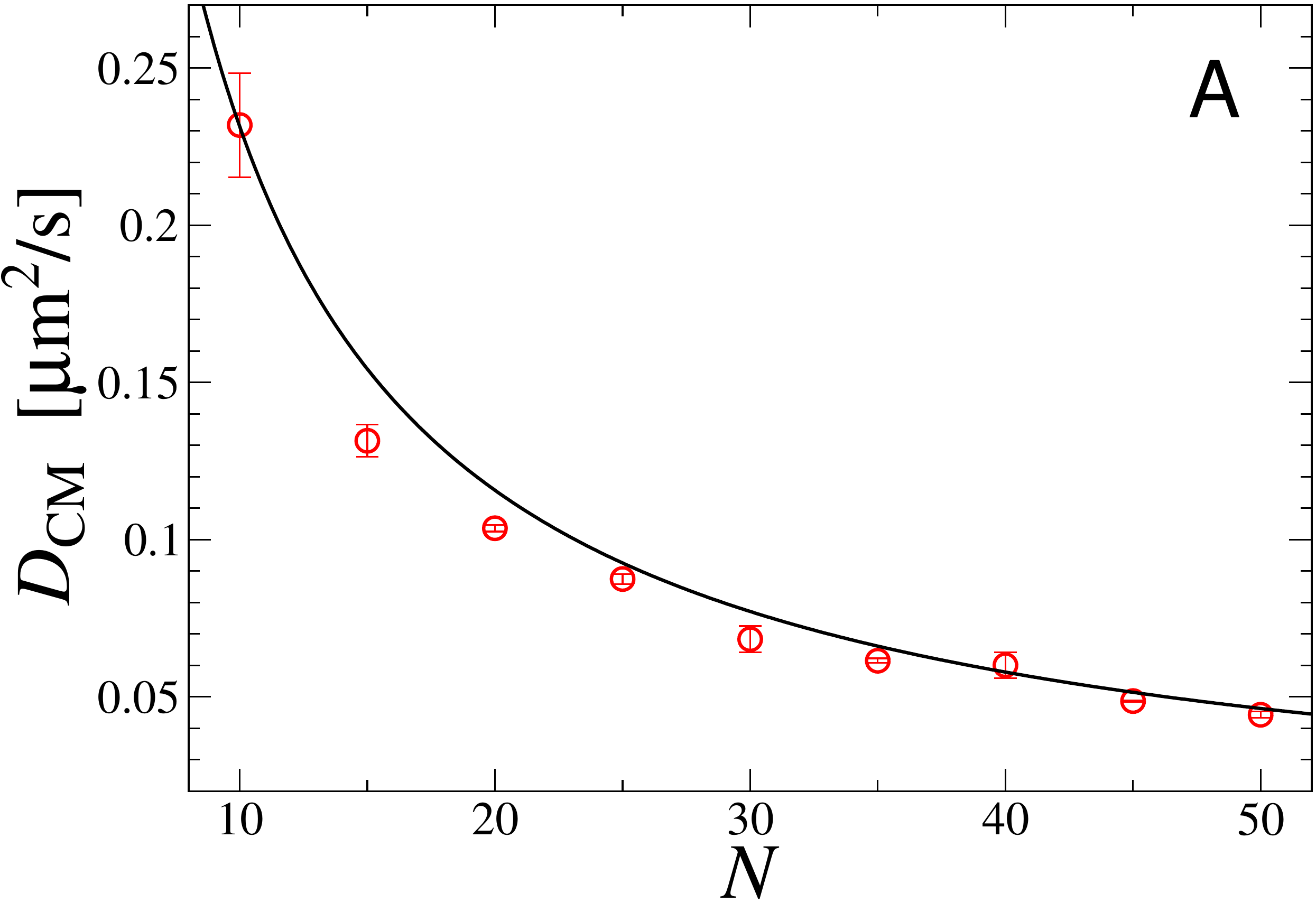}
\hspace{0.2cm}
\includegraphics[width=0.45\linewidth]{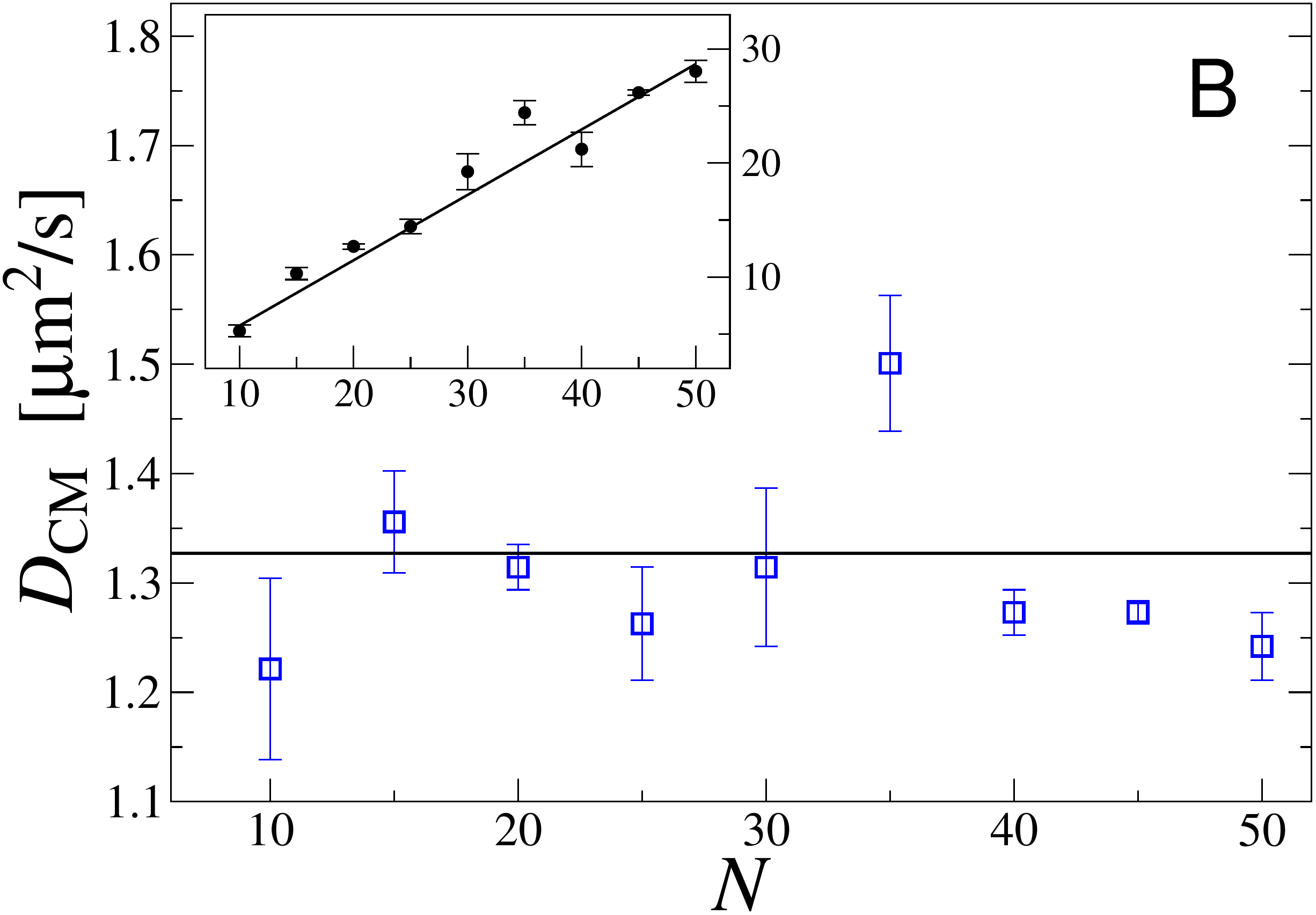}
\caption{Center-of-mass diffusion coefficient as a function particle number, obtained from simulations without (A) and with (B) flow-mediated correlations. The solid curve in Panel~A shows the theoretical prediction in the absence of correlations, $D_{\rm CM}^{\rm un}=D_0/N$ (Eq.~\ref{Dcm no corr}). The solid line in Panel~B shows the theoretical prediction in the presence of correlations, with $\Dcm$ equivalent to that of a rigid disk of radius $R_0$ (Eq.~\ref{Dcm final} for $t=0$). The inset shows the ratio between the coefficients with and without correlations. Parameter values are as in Fig.~\ref{fig:analytic}.}
\label{fig:Dcm}
\end{figure}

Figure~\ref{fig:Dcm}A shows the results for the CM diffusion coefficient without correlations. The numerical results follow Eq.~\ref{Dcm no corr} (solid line) nicely, demonstrating the expected decrease of $\Dcm$ with number of particles in the assembly. Panel~B shows the results in the presence of flow-mediated correlations. The CM diffusion coefficient is an order of magnitude larger than in Panel~A and remains roughly unchanged ($\sim 10\%$ variation) with increasing particle number. The numerical values are in quantitative agreement with $\Dcm(t=0)$ as obtained from Eq.~\ref{Dcm final} (solid line). The inset in Panel~B shows the ratio between the CM diffusion coefficients with and without the correlations, highlighting the strong effect of correlations and its agreement with the analytical theory.

\begin{figure}[hbt!]
\centering
\includegraphics[width=0.45\linewidth]{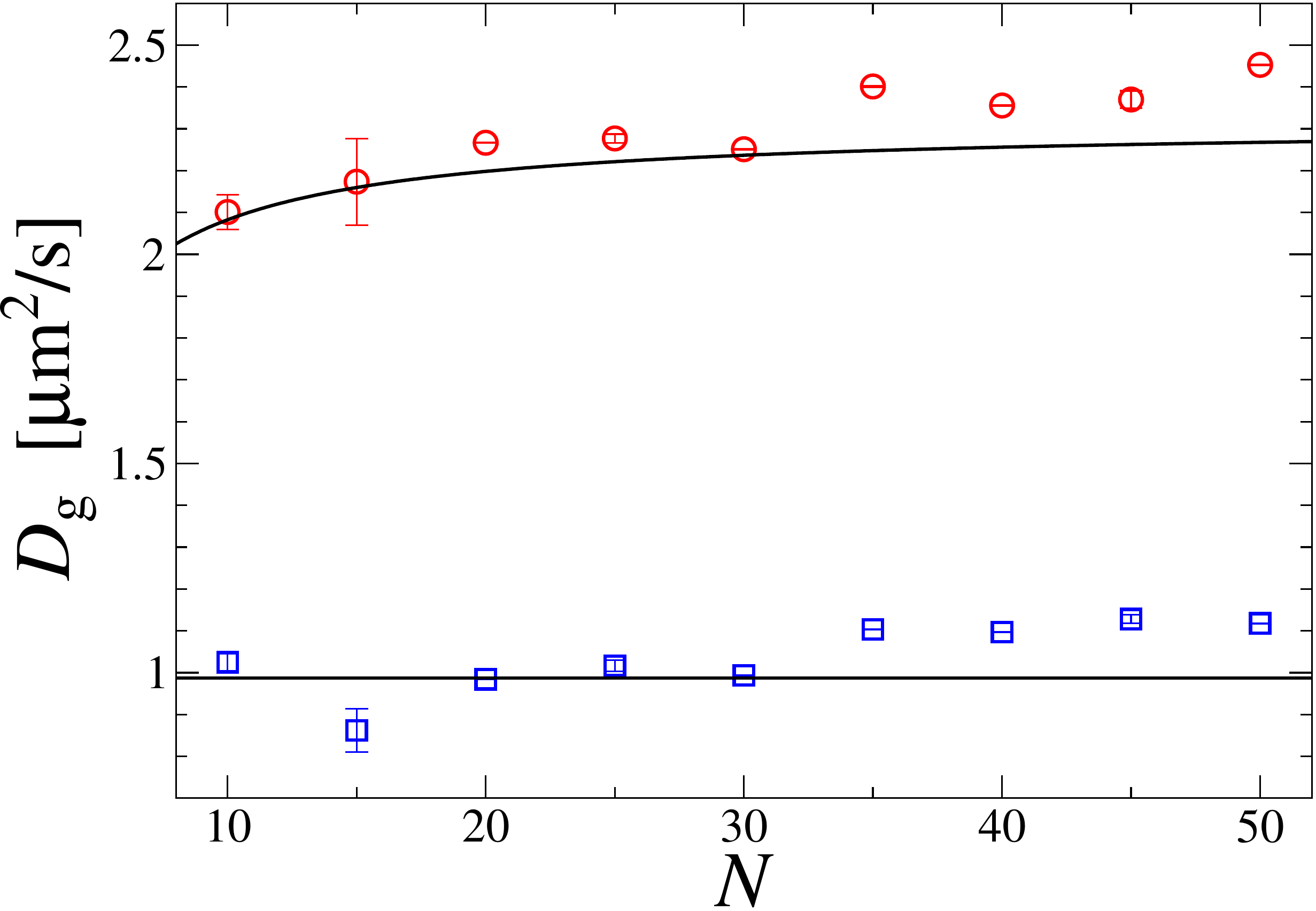}
\caption{Gyration diffusion coefficient as a function of particle number, obtained from simulations without (red circles) and with (blue squares) flow-induced correlations. This coefficient characterizes the diffusive dispersal of the assembly. The solid lines correspond to the analytical predictions (Eqs.~\ref{DRg as Dcm} and \ref{Dcm no corr} for the upper curve; Eqs.~\ref{DRg as Dcm} and \ref{Dcm final} for the lower line). Parameter values are as in Fig.~\ref{fig:analytic}.}
\label{fig:Drg}
\end{figure}

Figure~\ref{fig:Drg} presents the results for the gyration diffusion coefficient without and with correlations. The uncorrelated assembly is found to disperse twice as fast as the correlated one. The results are in quantitative agreement with the analytical predictions (Eq.~\ref{DRg as Dcm}). Comparing Figs.~\ref{fig:Drg} and \ref{fig:Dcm}B, we see that, for the selected (realistic) parameter values, the correlations make the gyration diffusion coefficient smaller than the CM one.

\begin{figure}[hbt!]
\centering
\includegraphics[width=0.45\linewidth]{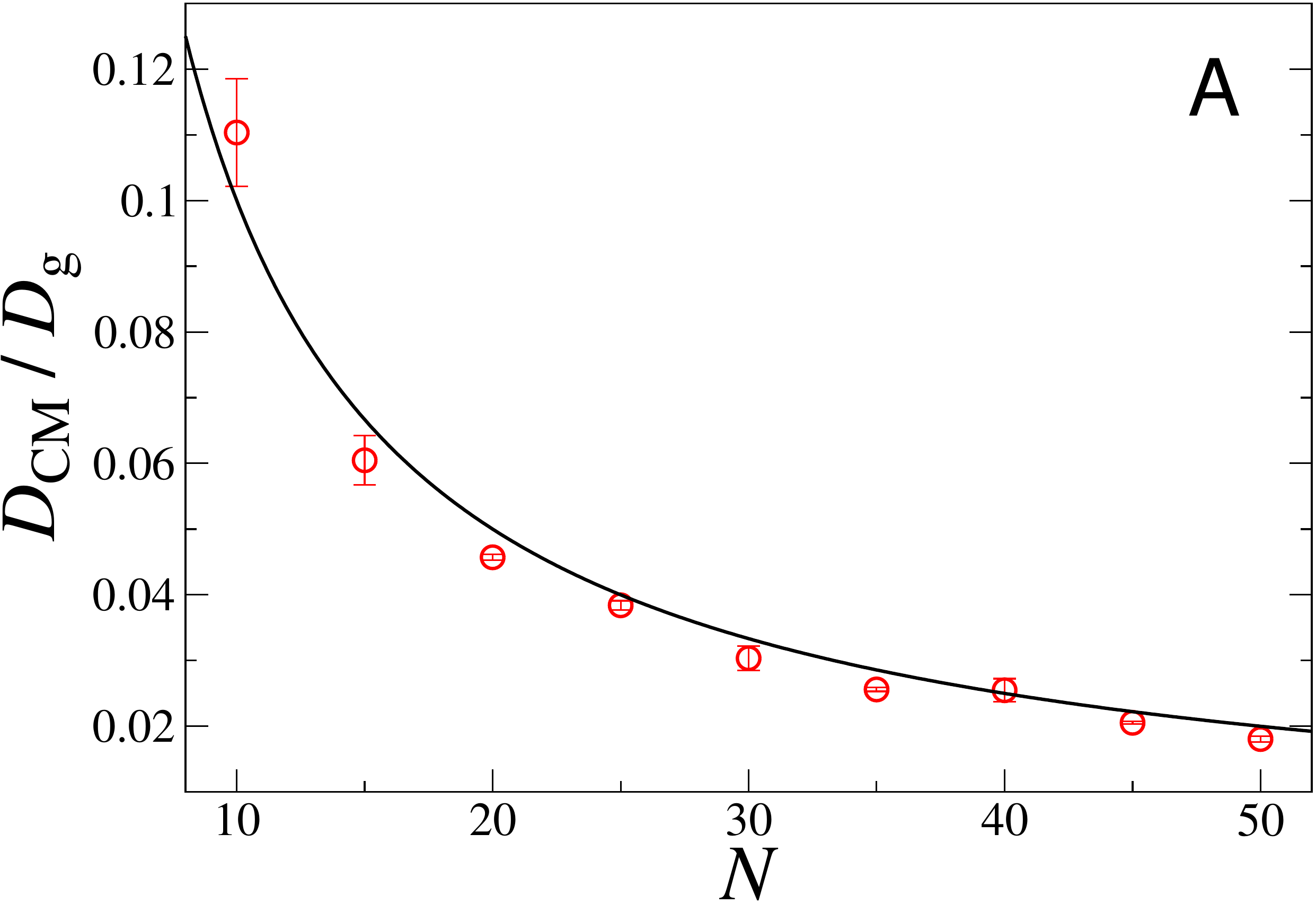}
\hspace{0.2cm}
\includegraphics[width=0.45\linewidth]{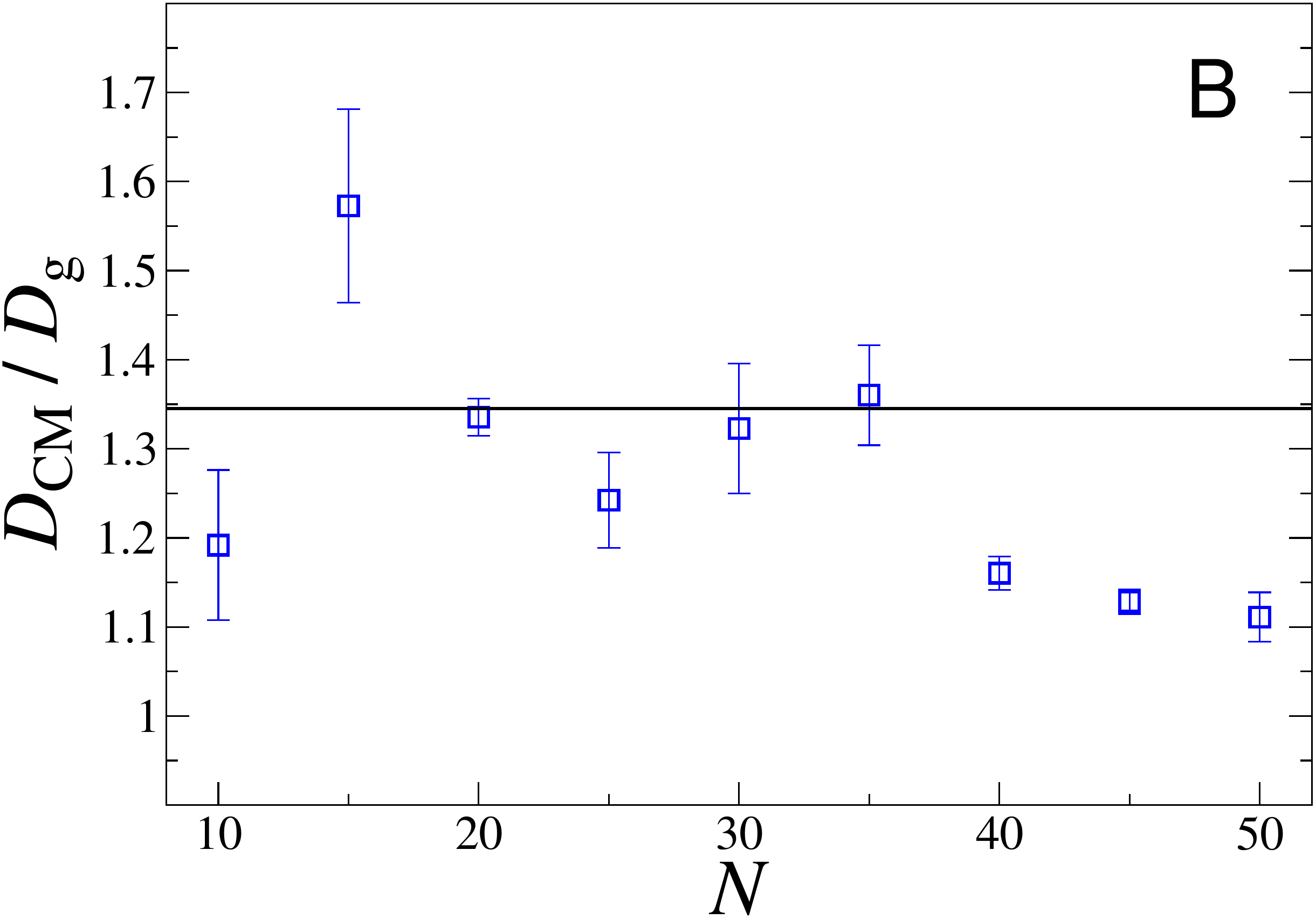}
\caption{Ratio of CM diffusion coefficient and gyration diffusion coefficient as a function of particle number, obtained from simulations without (A) and with (B) flow-induced correlations. The data are the same as in Figs.~\ref{fig:Dcm} and \ref{fig:Drg}. The solid lines correspond to the analytical predictions. Parameter values are as in Fig.~\ref{fig:analytic}.}
\label{fig:DcmDrg}
\end{figure}

Finally, in Fig.~\ref{fig:DcmDrg} we re-use the data of Figs.~\ref{fig:Dcm} and \ref{fig:Drg} to show the ratio between the CM diffusion coefficient and the gyration diffusion coefficient without and with flow-induced correlations. The larger $\Dcm$ (Fig.~\ref{fig:Dcm}), together with the smaller $\DR$ (Fig.~\ref{fig:Drg}), lead in the presence of correlations to a ratio that is larger by a factor of 10--50 than its values without correlations. For larger particle assemblies the ratio will increase even further.

\section*{Discussion}

The results that are most relevant to experiments are given in Eqs.~\ref{Dcm final} and~\ref{DR final} for the CM and gyration diffusion coefficients. They demonstrate how the two coefficients are comparable. This central result is a distinctive property of inclusions in fluid membranes, and differs qualitatively from the behaviors in other particle-laden fluids. As we have mentioned above, because of the weak logarithmic dependence on size, the time-dependent radius $\bar{R}\left(t\right)$ can be replaced without significant error by the initial domain radius $R_0$. This has been also demonstrated in our simulations. These two diffusion coefficients could be measured, for example, by fluorescence microscopy. 

On a more theoretical level, we obtained the exact solution for the time-dependent CM and gyration diffusion coefficients, along with their short-time and long-time asymptotes, for a Gaussian distribution of particles. See Eqs.~\ref{exact}--\ref{long} and Eqs.~\ref{DR exact}--\ref{DR long}, respectively. 
The time-dependence of the diffusion coefficients leads to a weak sub-diffusion of the CM (MSD growing in time as $t\ln \left(t^{-1}\right)$). This is a unusual example of scaled Brownian motion in a thermalized environment \cite{jeon2014scaled}. A few additional, theoretical findings are worth mentioning: 
(a) The correction to the correlation tensor (Eq.~\ref{Btensor}) due to the particles' finite size does not contribute to the diffusion coefficients. This implies that our fluid membrane is ultimately equivalent theoretically to a 2D fluid, which induces purely logarithmic correlations up to some system-dependent cutoff. (b) The diffusion coefficient explicitly characterizing the expansion of the domain, $D_{\rm exp}$, is found to be independent of any length scale in the system (see Eq.~\ref{exact}). (c) The gyration diffusion coefficient was found to be independence of the the SD length.

Our main analytical predictions have been confirmed by the simulations without any fitting parameters. At the same time, the simulations showed that the theoretical subtleties just described would be hard to resolve experimentally. 
At the long-times required to observe them, the assembly has already completely dispersed. For the same reason, $\Dcm$ is relevant only in the short-time limit, when the assembly in still small, requiring super-resolution microscopy. Numerically, observing the subtle sub-diffusion would require more extensive simulations than those presented here.

The most significant simplification in our theory is the absence of any direct interactions among the inclusions, as our purpose has been to isolate the essential effect of flow-mediated correlations. For example, short-range interactions at the perimeter of the domain would lead to effective line tension which has not been considered here. Neglecting direct interactions also leads to a gradient diffusion coefficient which is unaffected by the correlations. Its dependencies on the CM and gyration diffusive motions mutually cancel, leading to $\bar{D}=D_0$. More accurately it should depend on positional correlations, i.e., the structure factor \cite{pusey89}. Future studies should address the interplay between the flow-mediated correlations and molecular interactions. We note that at distances larger than a few nanometers (i.e., for sufficiently spread-out clusters) we expect the long-ranged flow-mediated effect to be dominant. Another strong assumption has been the consideration of a pristine membrane with unobstructed flows. In a real membrane, our results should still be valid over distances shorter than a certain cutoff length, replacing the SD length \cite{oppenheimer2010correlated,oppenheimer2011plane,chein2019flow}. 
Molecules which locally modify the membrane's fluidity \cite{marsh2010liquid,hodzic2012losartan}, such as cholesterol, should also influence the dispersal of domains.

There are several additional noteworthy approximations which may be relaxed in future studies. (a) We have assumed that the inclusions are much smaller than the inter-particle distances within the assembly. The interaction tensor that we have used (Eq.~\ref{Btensor}) contains the leading correction to this limit, which turns out to have no effect on the CM and gyration diffusion coefficients. Much denser assemblies where particles reach close proximity would require higher-order terms. (b) The assembly's evolving shape and motion were assumed to be isotropic, resulting in scalar diffusion coefficients. Asymmetric shapes would entail more complicated (tensorial) response, which might lead to interesting shape instabilities. (c) We have not included memory (viscoelastic) effects \cite{Camley11memory,komura2012anomalous}; 
our anomalous diffusion (sub-diffusion) arises from an instantaneous response to evolving configurations. Finally, (d) assuming a flat membrane, we have neglected membrane curvature and fluctuations. The effect of these factors on in-plane diffusion is subtle \cite{camley2010dynamic,henle2010hydrodynamics}, and their implications for domain dispersal require a comprehensive separate study.

\section*{Conclusion}

The numerical and analytical results as presented in Fig.~\ref{fig:DcmDrg} quantify our central observation\,---\,flow-induced correlations between membrane inclusions prolong the lifetime of membrane-embedded assemblies to a remarkable extent. This general observation may have profound implications for the stability of actual nano-domains and rafts in biological membranes. 
On top of the usually studied stages of molecular recruitment  into a functional membrane domain, we wish to highlight the ensuing stage of the domain's dispersal. Thus, the results obtained here may play a key role in more detailed modeling of various processes such as endo- and exocytosis, viral infection, and signaling.

\section*{Author Contributions}
The two authors contributed equally to the research and writing of the manuscript.

\section*{Acknowledgments}

We thank Frank Brown, Shigeyuki Komura, Naomi Oppenheimer, and Shlomi Reuveni for their helpful input. The research has been supported by the Israel Science Foundation under Grant No.~986/18.

\bibliography{membrane_domain}


\section*{Supplementary Material}

An online supplement to this article can be found by visiting BJ Online at \url{http://www.biophysj.org}.

\end{document}